\documentclass[journal]{IEEEtran}

\ifCLASSOPTIONcompsoc
  \usepackage[nocompress]{cite}
\else
  \usepackage{cite}
\fi

\usepackage{cite}
\usepackage{amsmath}
\usepackage{amssymb}
\usepackage{amsfonts}
\usepackage{graphicx}
\usepackage{textcomp}
\usepackage{xcolor}
\usepackage{multirow}
\usepackage{soul}
\usepackage{tikz}
\usepackage{algorithm}
\usepackage{algpseudocode}
\usepackage{colortbl}
\usepackage{bm}
\usepackage{float}
\usepackage{filecontents}

\usepackage{dblfloatfix}    

\newcommand{\cmmnt}[1]{}  

\def\BibTeX{{\rm B\kern-.05em{\sc i\kern-.025em b}\kern-.08em
    T\kern-.1667em\lower.7ex\hbox{E}\kern-.125emX}}

\newcommand\encircle[1]{%
\tikz[baseline=(X.base)] 
\node (X) [draw, scale=0.75, shape=circle, inner sep=0, fill=black, text=white, minimum size=0em] {\strut #1};}

\ifCLASSINFOpdf
\else
\fi
\begin{document}
%
\title{Neuro-Photonix: Enabling Near-Sensor Neuro-Symbolic AI Computing on Silicon Photonics Substrate}
%
%
%

\author{Deniz Najafi,~\IEEEmembership{Student Member,~IEEE,}
Hamza Errahmouni Barkam,~\IEEEmembership{Student Member,~IEEE,}
Mehrdad Morsali,~\IEEEmembership{Student Member,~IEEE,}
SungHeon Jeong,~\IEEEmembership{Student Member,~IEEE,}
Tamoghno Das,~\IEEEmembership{Student Member,~IEEE,}
Arman Roohi,~\IEEEmembership{Senior Member,~IEEE,}
Mahdi Nikdast,~\IEEEmembership{Senior Member,~IEEE,}\\
Mohsen Imani,~\IEEEmembership{Member,~IEEE,}
and~Shaahin~Angizi,~\IEEEmembership{Senior Member,~IEEE}
\thanks{D. Najafi, M. Morsali and S. Angizi are with the Department of Electrical and Computer Engineering, New Jersey Institute of Technology, Newark, NJ, USA. E-mail: dn339@njit.edu;mm2772@njit.edu;shaahin.angizi@njit.edu.}
\thanks{H. E. Barkam, S. Jeong, T. Das, and M. Imani are with the Department of Computer Science, University of California, Irvine, USA e-mail: herrahmo@uci.edu;sungheoj@uci.edu;tamoghnd@uci.edu;m.imani@uci.edu.}
\thanks{A. Roohi is with the Department of Electrical and Computer Engineering, University of Illinois Chicago, Chicago, IL, 60607. E-mail: aroohi@uic.edu.}
\thanks{M. Nikdast is with the Department of Electrical and Computer Engineering at Colorado State University, USA. E-mail: mahdi.nikdast@colostate.edu.}
\vspace{-2em}}

\markboth{}%
{Shell \MakeLowercase{\textit{et al.}}: Bare Demo of IEEEtran.cls for IEEE Journals}

\maketitle

\begin{abstract}
Neuro-symbolic Artificial Intelligence (AI) models, blending neural networks with symbolic AI, have facilitated transparent reasoning and context understanding without the need for explicit rule-based programming. However, implementing such models in the Internet of Things (IoT) sensor nodes presents hurdles due to computational constraints and intricacies. In this work, for the first time, we propose a near-sensor neuro-symbolic AI computing accelerator named Neuro-Photonix for vision applications. Neuro-photonix processes neural dynamic computations on analog data while inherently supporting granularity-controllable convolution operations through the efficient use of photonic devices. Additionally, the creation of an innovative, low-cost ADC that works seamlessly with photonic technology removes the necessity for costly ADCs. Moreover, Neuro-Photonix facilitates the generation of HyperDimensional (HD) vectors for HD-based symbolic AI computing. This approach allows the proposed design to substantially diminish the energy consumption and latency of conversion, transmission, and processing within the established cloud-centric architecture and recently designed accelerators. Our device-to-architecture results show that Neuro-Photonix achieves 30 GOPS/W and reduces power consumption by a factor of 20.8 and 4.1 on average on neural dynamics compared to ASIC baselines and photonic accelerators while preserving accuracy.
\end{abstract}


%
\IEEEpeerreviewmaketitle

\section{Introduction}
\IEEEPARstart{T}{he} Internet of Things (IoT) market is projected to reach \$1100 billion by 2025, supported by 75 billion connected devices in smart homes and cities. The vast data generated by these pervasive devices offers significant potential for intelligent IoT services, which can revolutionize modern life and drive AI advancements \cite{iot2022,tabrizchi2022tizbin}. However, current IoT devices rely heavily on cloud computing due to a lack of autonomous intelligence, leading to significant power demands and under-utilization of data \cite{song2022reconfigurable,xu2020macsen,angizi2023pisa}. Therefore, a shift towards a data-centric approach with more intelligent algorithms is imperative to tackle these challenges. This involves enabling IoT nodes to process data locally and generate interpretations for the cloud. In this way, there's a pressing need for transparent and interpretable IoT models to ensure efficiency and safety. Traditional models like Deep Neural Networks (DNNs) show promise but are critiqued for their 'black-box' nature, hindering abstract reasoning and context understanding~\cite{lu2023survey, helenli2024}. Neuro-symbolic AI presents a compelling solution, blending neural network pattern recognition with symbolic AI's logical reasoning capabilities. This fusion aims to provide transparent and logical solutions for a broad range of IoT tasks~\cite{garcez2020neurosymbolic}.

Neuro-symbolic AI  
is particularly valuable for reasoning tasks, enabling capabilities on sensing devices that were not feasible with traditional methods. Besides, such models aim to bolster model robustness and transparency and facilitate complex decision-making near the sensor. A significant challenge, however, lies in optimizing the interaction between hardware and algorithms. Current state-of-the-art solutions often require too much power, are large, or demand excessive computational resources \cite{wan2024h3dfact, tian_weakly_2022, mao_neuro-symbolic_2019, qu_probabilistic_2019}. Overcoming these obstacles is crucial for fully realizing the potential of neuro-symbolic AI, and this paper dives into the latest advancements and challenges in the hardware-algorithm optimization process, focusing on maximizing the benefits of neurosymbolic integration.

On the hardware side, advancements have primarily concentrated on augmenting CMOS image sensors to accelerate the processing of DNN workloads. One strategy involves integrating CMOS image sensors and processors onto a single chip, a concept known as Processing-Near-Sensor (PNS) \cite{najafi2024hybrid,carey2013100,najafi2024enabling,hsu20200,yamazaki20174,angizi2019mrima,tabrizchi2023nese,angizi2018cmp}. Another strategy entails embedding computation units within individual pixels, termed Processing-In-Sensor (PIS) \cite{xu2020macsen,xu2021senputing,tabrizchi2023appcip,angizi2023pisa,tabrizchi2024apris,tabrizchi2024vitsen,yun2024hypersense,morsali2023design}. The PIS framework handles pre-Analog-to-Digital Converter (pre-ADC) data processing before transmission to the on- or off-chip processor. Despite these advancements, particular challenges persist, including the energy consumption associated with ADC, Digital-to-Analog Converters (DAC), and sense amplifiers in PIS, thereby constraining the integration of all DNN layers within the pixel array \cite{el1999pixel,taheri2024ressen,morsali2024energy,song2022reconfigurable,abedin2022mr,ma2023leca,reidy2024hirise,morsali2023deep}.
The primary focus of existing research has been expediting the initial layer of DNNs and offloading the subsequent layers to a digital accelerator, primarily due to the limited resources of PIS systems. Consequently, the enduring issue of power-intensive peripherals and DAC/ADC units exists despite efforts to minimize their energy consumption for sensing and computation purposes \cite{choi2015energy,xu2020macsen,hsu2019ai,sunny2021crosslight}.
\textcolor{black}{RedEye \cite{likamwa2016redeye} utilizes charge-sharing tunable capacitors to perform convolution operations. While this approach achieves energy savings compared to CPU/GPU implementations at the cost of reduced accuracy, maintaining high computational accuracy significantly increases energy consumption, requiring up to 100$\times$ more energy per frame. Macsen \cite{xu2020macsen}, a PIS platform, processes the first convolutional layer of Binary Weight Neural Networks (BWNN) using a correlated double sampling procedure, enabling computation speeds of 1000 fps. However, it faces substantial challenges due to its large area overhead and high power consumption, primarily caused by the SRAM-based PIS methodology.}
Besides, the notable area overhead and power consumption associated with recent PNS/PIS units necessitate additional memory for storing intermediate data \cite{angizi2023pisa,tabrizchi2023appcip,song2022reconfigurable}. Moreover, the computational speed stemming from electronic systems operating at limited gigahertz frequencies cannot match the high speeds and extensive parallelism characteristic of optical systems with photo-detection rates surpassing 100GHz \cite{sunny2021robin,morsali2024oisa,sunny2021crosslight,cheng2020silicon,morsali2024lightator}. 

In recent years, several designs have focused on utilizing photonics to enhance computational speed. \cite{wang2018} developed a design employing photonics to implement basic logic gates. Additionally, \cite{liu2019holylight} introduced a nanophotonic accelerator for CNN implementations, incorporating shifter and adder components within the CNN. While these designs leverage silicon photonics to improve speed and parallelism, none have effectively implemented the MAC operation using silicon photonics to enhance efficiency in implementing neural networks.
Apart from that, few studies have attempted to incorporate neuro-symbolic methods into in/near-sensor computing environments, as detailed by Arrotta et al. in \cite{arrotta2024neuro}. However, these implementations are predominantly confined to relatively straightforward classification tasks. Conversely, Wan et al. \cite{wan2024h3dfact} proposed an advancement in accelerating visual reasoning applications through the factorization of holographic perceptual representations. While this approach promises significant acceleration, it necessitates a multi-tiered 3D memory hierarchy, which is currently impractical for in/near-sensor integration. This discrepancy underscores a critical gap in the literature, highlighting the need for more feasible integration strategies that can accommodate complex computational frameworks in constrained environments.

In this work, we propose a sensor-aware neuro-symbolic AI framework that synergistically combines the strengths of deep learning and symbolic AI on a Silicon Photonics substrate.
Unlike other traditional neuro-vector-symbolic architectures (NVSA), which depend on conventional electronic hardware and use hybrid analog-digital acceleration to optimize neural network quantization, necessitating expensive ADCs and resulting in high power consumption, the proposed Neuro-Photonix architecture utilizes a silicon photonics substrate. This approach leverages photonic devices for neural dynamic computation, thereby minimizing the need for costly conventional ADCs, significantly reducing energy consumption, and achieving faster processing speeds.
The main contributions of the work are as follows. (1) We introduce \textit{Neuro-Photonix}, a cutting-edge optical near-sensor accelerator tailored for Neuro-Symbolic AI. It offers high performance and energy efficiency, capable of handling the complete processing of neural dynamics and symbolic AI hyper-vector generation; (2) We design a neuro-symbolic AI model compatible with the hardware in terms of bit-precision on the input and the weights while maintaining state-of-the-art performance on complex reasoning tasks such as the RAVEN dataset \cite{raven2019}; and (3) We present a comprehensive study of the different hyperparameters of the algorithm and the design space of the hardware and its impact on performance markers such as accuracy, energy consumption, and latency. All these contributions enable the first-of-its-kind near-sensor neuro-symbolic AI system to perform reasoning tasks near the sensor realistically.

The rest of the paper is structured as follows: Section II delves into current advancements in Silicon Photonics acceleration mechanisms and Neuro-Symbolic AI. Section III outlines the proposed Neuro-Photonix's circuit-to-architecture co-design methodology. Section IV details the innovative correlated and photonics-optimized hardware mapping technique for the Multiply-Accumulate (MAC) engine and encoder. Section V describes the bottom-up evaluation framework and presents a thorough analysis of the evaluation results and comparison. Finally, Section VI concludes the paper.

\vspace{-0.8em}

\section{Background and Related Work}
\textbf{Silicon Photonics Acceleration.}
Silicon-photonic-based accelerators, noted for their higher operational bandwidth compared to electronic ones and their resolution of fan-in/fan-out issues, are preferred for improving DNN and machine vision tasks \cite{sunny2021arxon,sunny2021crosslight,liu2019holylight,zokaee2020lightbulb}. These devices are categorized into coherent and non-coherent designs. Coherent designs use a single wavelength with the weight and activation parameters encoded in the amplitude, phase, or polarization of the optical signal \cite{zhao2019hardware}. Non-coherent designs utilize multiple wavelengths to perform parallel computations, encoding parameters in the signal's amplitude \cite{sunny2021crosslight,sunny2021robin}. Micro-Ring resonators (MRs) are employed to manage specific wavelengths and can be dynamically adjusted using mechanisms like microheaters or PIN junctions to selectively interact with designated wavelengths \cite{sunny2021crosslight,sunny2021robin}.

\begin{figure}[t] 
\centering
\includegraphics [width=0.94\linewidth,]{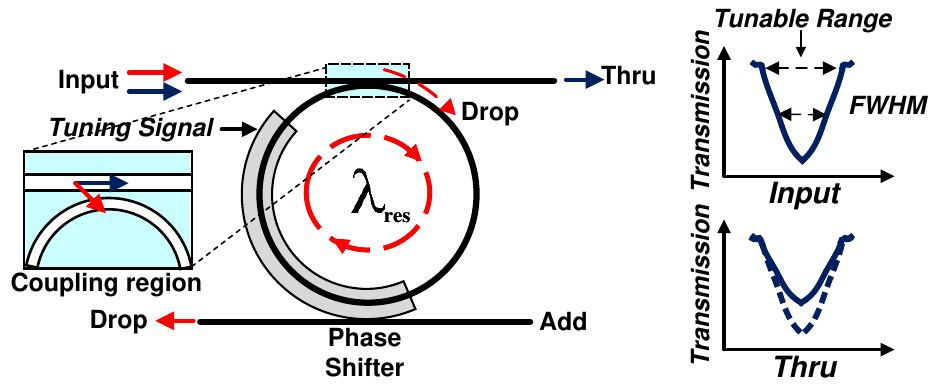}
\vspace{-1.2em}
\caption{MR input and through ports’ spectra after imprinting a parameter (using tuning signal). By adjusting the MR's resonant wavelength ($\lambda_{res}$) using the phase shifter, part of the input signal drops into the ring (through the coupling region) towards the drop port while the remaining propagates towards the through port.} 
\vspace{-1.0em}
\label{mr}
\end{figure}

In non-coherent systems, fine adjustments of MRs, effectively serving as weights, modulate the intensity of incoming light at specific wavelengths,  \cite{sunny2021crosslight,sunny2021robin}. MRs are essential for encoding weight and enhancing MAC operations by modifying the transmission spectrum of light. This modification is achieved by tuning the MRs' resonant wavelengths to overlap with the input light's wavelength, thus embedding the desired parameters into the transmission spectrum (as shown in Fig. \ref{mr}). The resonant wavelength is calculated with the formula $\lambda_{res}=\frac{{n_{eff}}\times L}{m}$, where $n_{eff}$ is the effective refractive index, 
$L$ the MR's circumference, and 
$m$ the resonant mode order \cite{bogaerts2012silicon}.
Prior research has delved into the enhancement of DNNs by leveraging both coherent and non-coherent photonic technologies. One notable proposal is LightBulb \cite{zokaee2020lightbulb}, a specialized accelerator for fully binarized Convolutional Neural Networks (CNNs) that employs photonic XNOR operations and popcounts instead of traditional floating-point MAC operations. While LightBulb achieves lower computation latency and requires less memory storage, its design is hampered by increased power consumption due to the extensive use of ADCs. Robin \cite{sunny2021robin} and CrossLight \cite{sunny2021crosslight} present low-bit-width weight-input CNN accelerators that require adjustment of both activation and weight values within the MRs, focusing solely on convolution layer operations similar to other designs. 
Another design outlined in \cite{sunny2022silicon} offers a CNN accelerator that supports mixed-precision weight-input by employing both wavelength-division multiplexing and time-division multiplexing within a non-coherent silicon photonic framework. Nonetheless, the continued reliance on DACs and ADCs as interlayer conversion mechanisms significantly increases the architectural footprint and power consumption. HolyLight \cite{liu2019holylight}, a nanophotonic accelerator, seeks to improve CNN inference throughput using MR-based adders and shifters instead of ADCs. Yet, the extensive application of MRs for encoding both activation and weight parameters leads to increased delays and power usage and diminishes the system’s adaptability for various DNN applications.

\begin{figure}[t!]
    \centering
    \vspace{-1mm}
    \includegraphics[width=1\columnwidth]{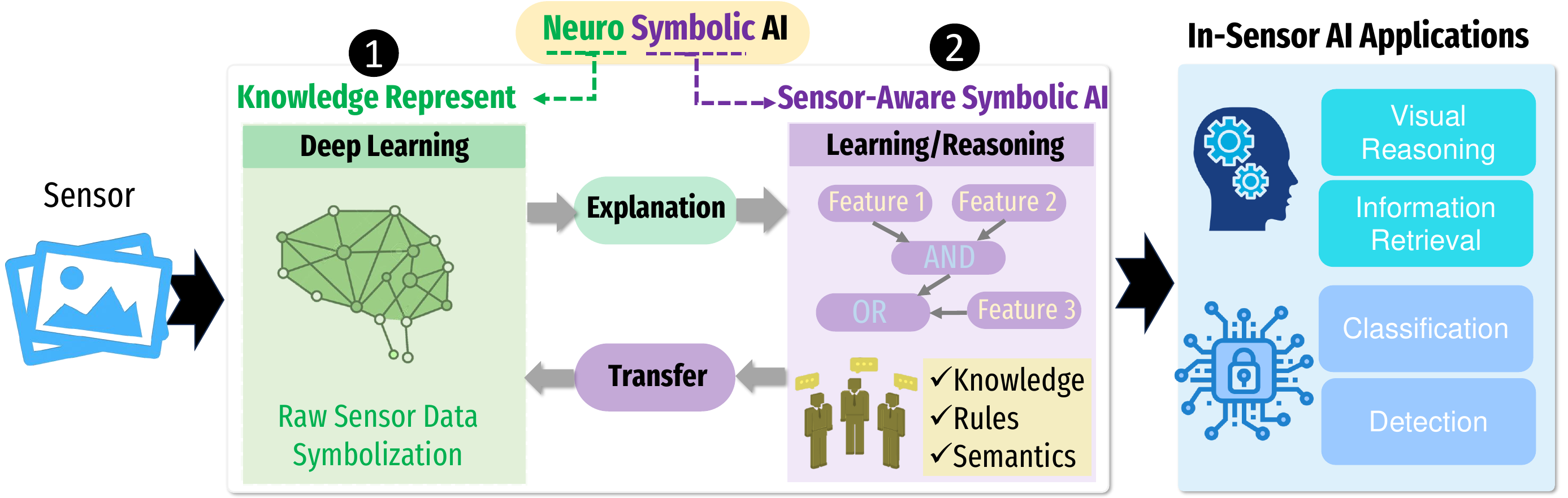}
    \vspace{-2em}
    \caption{Neuro-symbolic AI framework consisting of neural dynamic and symbolic AI models. Neural network models are responsible for information embedding, and symbolic models build up knowledge for symbolic decision-making.}
    \label{neuro}
   \vspace{-1.9em}
\end{figure}

\textbf{Neuro-Symbolic AI.} In the realm of machine learning for near-sensor IoT, we face unique challenges that differ from traditional edge computing scenarios. Notably, near-sensor AI algorithms are required to effectively process highly noisy analog data. For these algorithms to be viable, they must be highly efficient, memory-centric, robust, and transparent in their computations. These requirements make the deployment of reasoning capabilities directly within sensors particularly demanding. Symbolic AI is recognized for its strong performance in reasoning tasks, robustness against noisy conditions, and ability to make transparent decisions. However, our initial experiments have exposed a critical vulnerability; symbolic AI models lose a significant amount of their robustness when subjected to post-training quantization. This issue makes it challenging to deploy fully trained, advanced models within the limited bit precision environments typical of near-sensor devices. Additionally, symbolic AI struggles to extract information directly from raw sensor data. Given these constraints, there is a clear need for a hybrid approach.

By combining neural networks, which excel at extracting information, with symbolic AI, known for its decision-making strength, we can integrate what is known as Neuro-Symbolic AI into IoT devices. This hybrid approach aims to leverage the strengths of both domains to address the distinctive challenges faced by in-/near-sensor AI. As illustrated in Fig.~\ref{neuro}, neuro-symbolic AI algorithms have been introduced as a hybrid solution that combines deep learning capabilities with symbolic AI's transparency.
Computationally, the neural dynamic part \encircle{1} can be realized using various DNN models. However, the symbolic AI  part \encircle{2} requires programmability to support various learning and reasoning algorithms that can be achieved using HyperDimensional Computing (HDC).
HDC has demonstrated proficiency in addressing diverse tasks and datasets, establishing itself as a robust framework well-suited for applications requiring lightweight and highly efficient training \cite{hdc_review}. A noteworthy attribute contributing to HDC's popularity is its model weights resilience, specifically the dimensions of its Hyper-Vectors (HVs), as evidenced in prior works \cite{robust_1}. Furthermore, HDC exhibits resilience in handling non-idealities from hardware implementations, as corroborated by recent literature \cite{karunaratne2020memory}. \vspace{-1em}

\section{Neuro-Photonix}
In this work, we propose the Neuro-Photonix as a high-performance, energy-efficient, and versatile PNS accelerator to support neural symbolic AI at the edge. The distinct hardware needs for neural processing and symbolic reasoning present avenues for tailored optimizations. Neural dynamic components within such models can function efficiently at reduced precision through analog near-sensor computation on a silicon photonics substrate. This enables swift single-cycle MAC operations and exceptionally high energy efficiency while yielding dependable latent representations. Conversely, symbolic reasoning necessitates higher precision to employ symbols and derive precise decisions effectively.
In pursuit of achieving an effective neural symbolic AI architecture, we introduce a novel design aimed at enhancing precision through integrating HDC following the neural dynamic Component. Furthermore, according to previous studies, the approach of initially training neural networks and subsequently deriving an HDC model from the trained network led to a significant improvement in the accuracy of the overall design \cite{arxivma}. 

\subsection{Operational Flow} 
\begin{figure}[t]
\centering
\includegraphics [width=0.99\linewidth]{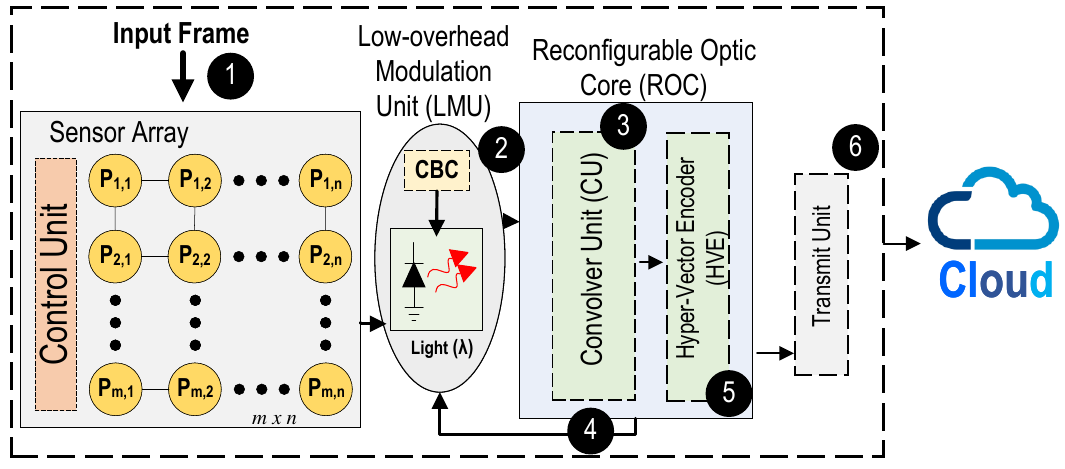}
\vspace{-1.5em}
\caption{High-level operational flow of the proposed Neuro-Photonix architecture.} 
\vspace{-1.1em}
\label{highlevel}
\end{figure}

Fig. \ref{highlevel} shows the high-level operational flow of the proposed Neuro-Photonix architecture. This architecture comprises a m$\times$n sensor array, a novel Low-overhead Modulation Unit (LMU), and a Reconfigurable Optical Core (ROC) as the main components. The operational sequence commences with the global shutter image sensor capturing the input frame (step \encircle{1}), followed by processing through the proposed LMU. 
In step \encircle{2}, the LMU transforms the sensor array data into a light beam characterized by varying wavelength ($\lambda$) and intensity, which is then transmitted to the ROC via the generated waveguide. Within the LMU, a novel Comparator-based Converter (CBC) is employed to convert the analog data into digital form efficiently and with minimal overhead. The ROC is responsible for running energy-efficient and low-precision dynamic neural networks, followed by converting the outcomes of the neural network into HV. This HV then serves as the foundational element of HDC. To accomplish this, in step \encircle{3}, each layer of a DNN is sequentially processed within the ROC. The ROC incorporates a Convolver unit (CU) tasked with handling the various layers of the DNN. The output of CU is then fed back to the LMU in step \encircle{4} as input to execute the next layer in the DNN. This iterative process continues to implement the entire network.In step \encircle{5}, the output of the DNN is encoded into an HV using the Hyper-vector Encoder (HVE). This involves reconfiguring the weights of the DNN within the ROC into high-dimensional encoding matrix values. After the creation of the HV, these vectors are transferred to the cloud for further processing (step \encircle{6}). The key advantage of the ROC lies in its simplicity in managing weight data, which only requires mapping onto MRs. It is worth noting that, in contrast to previous designs, the activation values are directly modulated into the core’s input using Vertical-Cavity Surface Emitting Lasers (VCSEL). This is achieved by adjusting the driving currents of the VCSELs, streamlining the process significantly. \vspace{-1em}

\begin{figure*} 
\centering
\includegraphics [width=0.84\linewidth]{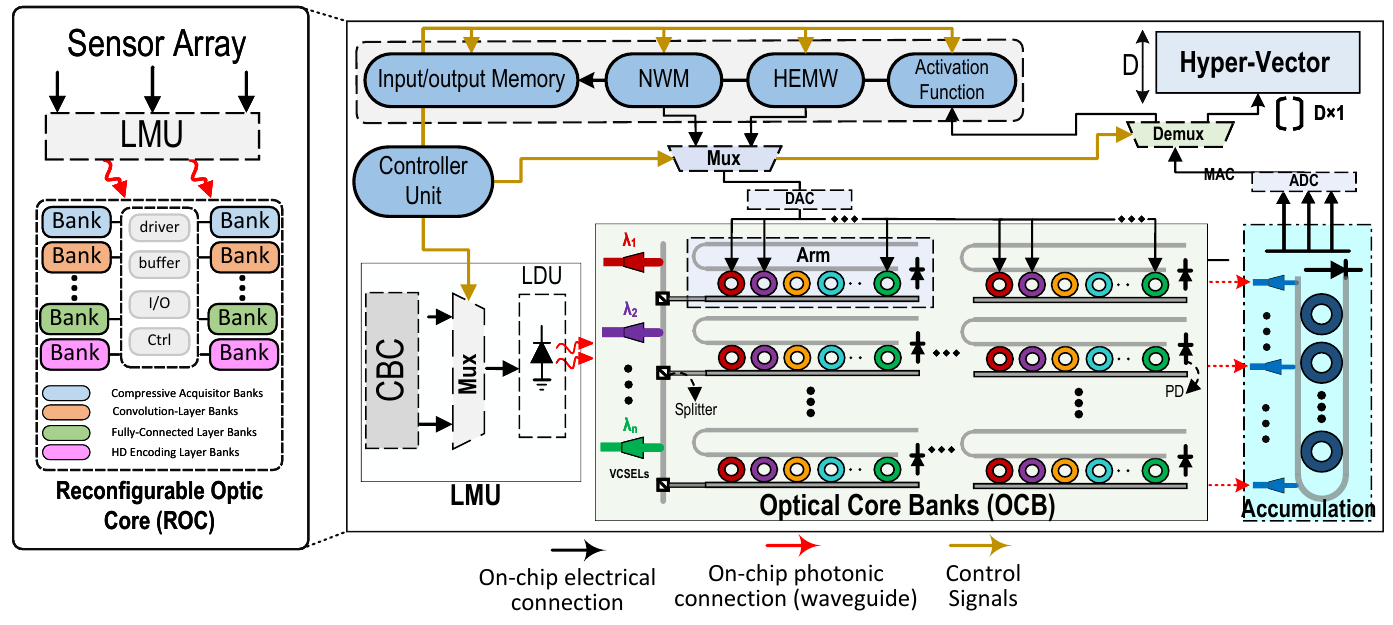}
\vspace{-1.7em}
\caption{Neuro-Photonix architecture consisting of a sensor array
and the optical core.} 
\vspace{-1em}
\label{arch}
\end{figure*}

\subsection{Architecture Design}
The main architecture of the proposed Neuro-Photonix system is depicted in Fig. \ref{arch}, comprising the LMU and the proposed ROC. The LMU is comprised of a CBC followed by a Light Driver Unit (LDU). This block is responsible for emitting a laser with varying intensities and wavelengths. Subsequently, this light enters into the ROC architecture. The ROC comprises Optical Core Banks (OCB) consisting of MRs followed by the Accumulation unit, which is essential for executing neural network layers and conducting encoding calculations crucial for HDC. These operations are conducted based on the MRs within the banks. 

In this architectural setup, the input light data first enters the OCB. Within the OCB, each neural network layer is executed by enabling the multiplication operation in its banks, leveraging the MRs for computation. Then, by using the Accumulation unit, the complete MAC operation needed for each layer of the neural network is performed. The output from each layer is subsequently fed back to the LMU as input for the next layer in the network. This feedback loop enables iterative processing of the input data through successive neural network layers. It is worth noting that the primary advantage of this architecture lies in mapping neural network weights to MRs, coupled with the activation of the layer results within the input light through the LDU. The weight values of the neural network are sourced from the Neural Weight Memory (NWM), facilitating efficient and integrated processing of neural network computations. This approach optimizes the utilization of optical components for weight mapping and activation, enhancing the overall performance and flexibility of the neural network implementation. 

After implementing the entire network, the outputs must be multiplied with the encoding matrix to generate the HV. This multiplication operation, essential for calculating the HV, is efficiently executed using MAC operations within the OCB. To this end, the weights stored in the MRs undergo reconfiguration using  Hyper-vector Encoding Memory Weights (HEMW). This process entails transferring each value of the encoding matrix, stored in the HEMW, to the MR elements. This step is crucial for encoding the neural network's output into an HV representation to improve robustness. By reconfiguring the weights stored in the MRs using the HEMW, the architecture enables the transformation of neural network outputs into hyper-dimensional vectors. \textcolor{black}{We have adopted a modular design approach, allowing individual components to be adapted or replaced based on specific application requirements. This modularity ensures flexibility and facilitates the integration of new technologies while maintaining system functionality. Furthermore, we have aligned our design with emerging standards in photonic and electronic integration to enhance compatibility with existing platforms and ensure long-term scalability.} A detailed explanation of the various components of the Neuro-Photonix architecture is elaborated in the following:

\textbf{(a) Low-overhead Modulation Unit (LMU).} The input image is acquired using a global-shutter RGB image sensor equipped with a photodiode in each pixel. The intensity of incident light generates a voltage drop in the photodiode, which is then transferred to the LMU. Within the LMU, the analog output from the pixels is directed to the CBC, which converts the analog data to 4-bit digital data. Notably, the utilization of CBC eliminates the need for power-intensive and area-consuming conventional ADC. The resulting digital data is subsequently transmitted to the LDU. This unit is designed to convert electrical input into light output at a specific wavelength and intensity using VCSEL. The input to the LDU originates either from the CBC unit or the preceding layer of the neural network.
Fig. 5 illustrates the detailed design of the LMU. The CBC unit comprises 15 comparators that compare the analog output of the photodiode with reference voltages ($V_{ref}$) spanning the range of pixel output voltages. Each comparator in the CBC unit produces either a `0' or a `1' based on the level of the input voltage it receives. This arrangement allows the CBC to generate 16 distinct outputs, representing 4-bit digital data (Fig. 5(a)). The generated data is then fed to the LDU to produce light with different wavelengths and intensities therefore, this design eliminates the need for a high-power, accurate ADC. It is important to note that a Multiplexer (Mux) is required between the CBC and LDU to select either the data coming from the pixel array or the output of the previous neural network layer (calculated in OCB) for transmission to the LDU.
The LDU consists of 15 transistors that are controlled by the output of the Mux, as shown in Fig. 5(b). The number of activated transistors in the LDU determines the current flow, resulting in higher light-intensity generation. It is noteworthy that conventional ADCs combine latch memory and an encoder with comparators to generate the appropriate output \cite{ADC2016}. In contrast, this design directly uses the outputs of 15 comparators to produce the proper light with the appropriate intensity. Therefore, the proposed design reduces overall power consumption and enhances efficiency by omitting the latch memory and encoder.

\textbf{(b) Optical Core Banks (OCB).}
This unit comprises multiple MR banks, visually represented with color-coding as illustrated in Fig. \ref{arch}. The various banks of OCB consist of a compressive acquisition, convolutional layer, fully-connected layer, and HD encoding layer. These components perform different DNNs and transform their outputs into HV by adjusting weight parameters and conducting matrix-vector multiplication operations as necessary.

\textit{1. Neural Dynamic Implementation.}
The OCB contains MR banks associated with specific weight values and partitioned into arms. These MRs modulate the intensity of incoming light based on their assigned weight values, selectively affecting light of the same wavelength as the MR. Implementing neural networks requires careful consideration of the MAC operation implementation, which is essential for efficient computation and performance. This operation involves multiplying each input by its corresponding weight and then accumulating these products to generate the output activations.
To this end, in OCB, the input light data, modulated to a specific intensity and wavelength within the LDU to represent required activation values, is transmitted through an arm containing the MRs that house the weight values (see Fig. \ref{arch}). As the light passes through the arm, each MR influences the intensity of light at the wavelength corresponding to that particular MR to result in W$\times$A which is known as partial product.
 At the end of each arm, a photonic detector is positioned to sum up extracted partial products and convert the processed light passing through the arm into a voltage signal.
 This process mirrors the MAC operation, which is essential for implementing various layers of neural networks. 
It's important to note that the number of MAC operations in each arm is limited by the number of MRs within that arm. Therefore, implementing large layers that require a large kernel size can be achieved by segmenting the required MAC operations, each fitting within an arm's capacity. In the accumulation phase shown in Fig. \ref{arch}, the summation of the MAC operations from each segment is computed to produce the overall result of the MAC operation. This involves adding together the individual products obtained from the MAC operations performed within each segment, resulting in the final output. 
It's important to note that this segmentation approach enables the implementation of complex neural network layers within the limitations of the hardware architecture.
Furthermore, due to the mentioned limitations, an electronic memory (NWM) is necessary to store the weights of the entire neural network. Another component in the electronic domain is the activation function that has been implemented at the end of each layer. During the implementation of each layer, the relevant weights are transferred from the NWM into the MR banks to execute the necessary MAC operations. This process ensures that the neural network computations can be carried out effectively within the hardware constraints. In addition to storing weights, additional memory is needed to store the output of each layer as activations, which serve as inputs for the subsequent layer in the neural network. \textcolor{black}{It is noteworthy that there are implementations of non-linear activation functions utilizing optical designs to ensure uniformity across architectures. For instance, the work presented in \cite{afifi2023tron} implements the GELU function. However, using memristors alongside MRs introduces design complexity and integration challenges. An MZI-based ReLU-like activation function has been implemented in \cite{williamson2019reprogrammable}. Nevertheless, employing MZI-based activation functions in MR-based architectures imposes additional design overheads due to the higher power consumption of MZIs compared to MRs and having different integration processes.}

\textit{2.	Symbolic AI's Hyper Vector Implementation.}
Upon implementing the DNN, it becomes necessary to map the outputs to HV representations to enhance efficiency \cite{arxivma}. This mapping is achieved by multiplying the DNN's output data into the encoding HV matrix, whose values are stored in the HEMW. These stored values are utilized to reconfigure the MRs within the banks, thereby constructing the encoding matrix within the OCB.
The encoding matrix is pivotal in mapping the network's output to a high-dimensional vector with a dimensionality of $D$, thereby enhancing precision and effectiveness in data representation. 
To create the HV, matrix-vector multiplication is employed to multiply the output data of the DNN with the encoding matrix. This process is initiated by inputting the DNN outputs, which are converted into light using the LDU, into the OCB containing the encoding matrix. 
The results of these operations yield an HV representation with a dimensionality of $D$. Fig. \ref{arch} visualizes this process, highlighting the flow and transformation of data as it progresses through the encoding and mapping stages to produce the HV representation ultimately.
\vspace{-0.8em}

\begin{figure}
\begin{center}\vspace{-1em}
\begin{tabular}{ll}
\includegraphics [width=0.48\linewidth]{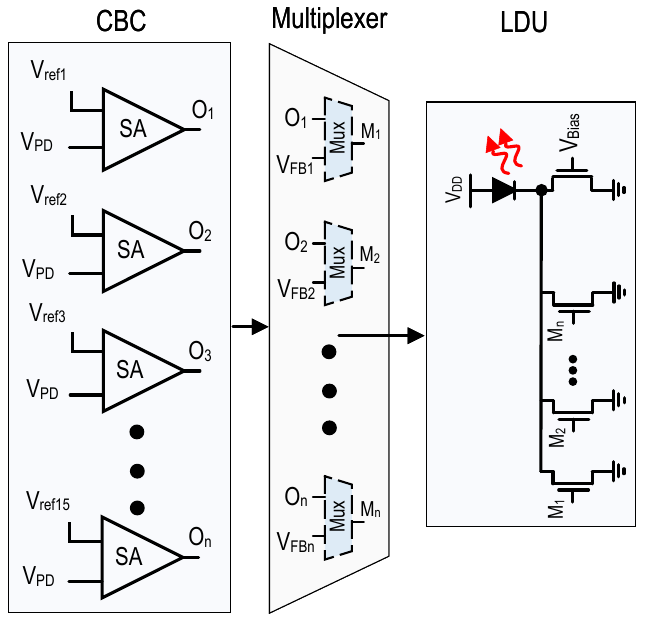} &
\includegraphics [width=0.54\linewidth]{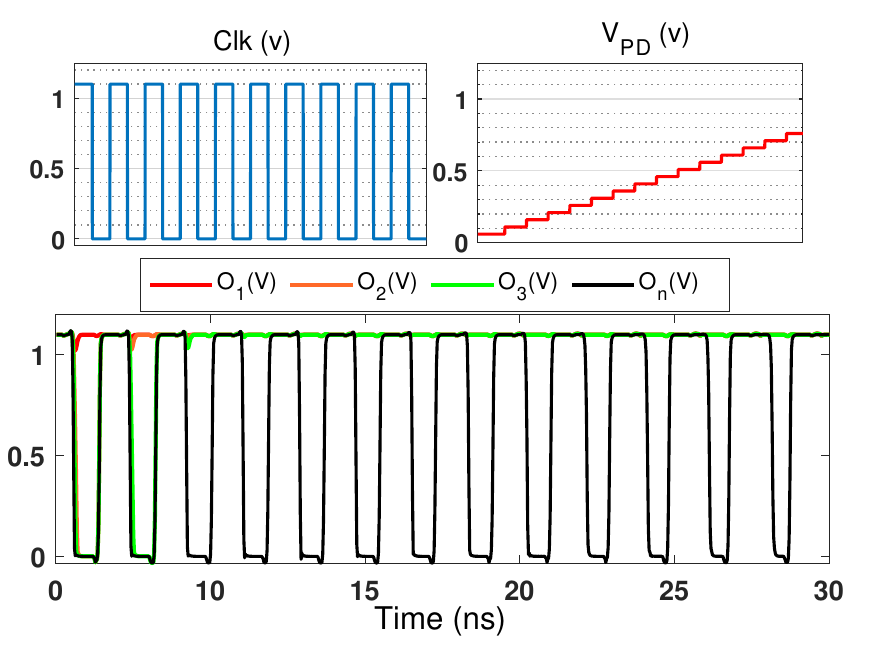}\\
 \end{tabular} 
 \vspace{-1em} (a) \hspace{7em} (b) \hspace{10em} (c)
\caption{Schematic of (a) Comparator-based Convertor (CBC) and (b) Light Driver Unit (LDU), (c) Sample waveforms of CBC input from the pixel and respective outputs.} \vspace{-2.5em}
\label{comparator}
\end{center}
\end{figure}


\section{Hardware Mapping}
Neuro-Photonix employs a correlated hardware mapping technique to effectively manage the computational demands of both neural dynamics and symbolic AI HV generation components. The goal is to evenly distribute the workload, thereby boosting processing speed while taking into account the physical constraints of the silicon photonics substrate and the power and area requirements for edge acceleration. We explore the design space of each component separately as follows.

\subsection{Photonics-Friendly MAC Engine}
\begin{figure}[t] 
\centering
\includegraphics [width=0.89\linewidth]{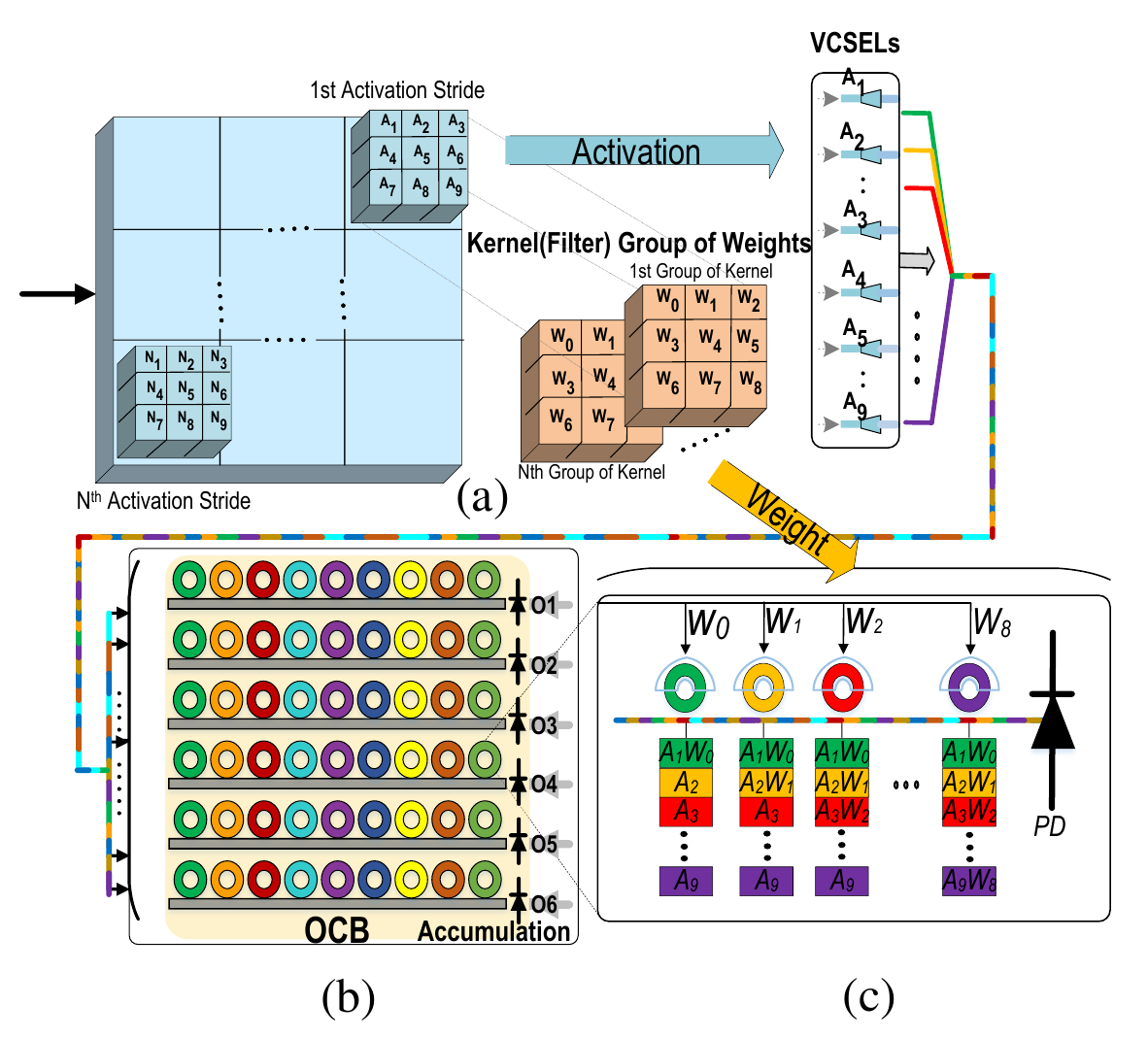}
\vspace{-1.7em}
\caption{(a) Activation modulation, (b) Hardware mapping for 6 Strides, (c) Implementing a 3$\times$3 kernel in an arm.}
\vspace{-1.5em}
\label{mapping}
\end{figure}

In the proposed architecture, the OCB is responsible for executing various layers of DNNs, including the convolutional layer and fully connected layer. To compute these layers, the weights of each DNN layer stored in the NWM need to be transmitted to the MRs situated within the OCB banks (refer to Fig. \ref{mapping}(a)). It is crucial to note that the NWM capacity is determined based on the maximum number of weights needed to accommodate the largest neural network layers. In this design, ResNet-18 is utilized as the benchmark for the maximum network size, thus a 5.5MB NWM is adequate to store the necessary 4-bit precision weights in the proposed design. 
In the initial step, the weights of each DNN layer are transferred from the NWM to the MRs.
In the proposed design, each bank row, an arm, contains 9 MRs. This configuration is chosen due to the common use of the 3$\times$3 kernel size in most DNN implementations. Additionally, to support kernel sizes of 5$\times$5 and 7$\times$7, each bank in the OCB is designed with a total of 54 MRs, which means each bank consists of 6 arms. This total of 54 MRs per bank ensures that even the maximum kernel size of 7$\times$7 can be processed in a single clock cycle. The proposed design allows for efficient utilization of each DNN layer by implementing the MAC operation in a single cycle.
For example, to implement a layer of DNN using a 3$\times$3 kernel size, each bank can execute 6 strides (O1-O6), with each arm responsible for one stride, as illustrated in Fig. 6(b). This process involves multiplying the input light data (A1-A9) by the weights stored in the MRs of an arm (W0-W8), as shown in Fig.\ref{mapping}(c). The weights, received in groups of nine from NWM, are stored in each arm to perform each kernel multiplication. A Photo Detector (PD) performs the summation operation, allowing the MAC result to be directly sent out without requiring an Accumulation unit at the end of each arm. This component is deactivated (shown in gray in Fig.\ref{mapping}(b)) during the implementation of the 3$\times$3 kernel.
Moreover, to implement larger kernel sizes such as 5$\times$5, 25 MRs are required for the multiplication process. This requires three arms in a bank, totaling 27 MRs, to execute one stride effectively. Since each arm cannot directly sum all 25 multiplication elements, an additional summation (Accumulation) is necessary. It should be noted that a single bank in the OCB can execute 2 strides for a 5$\times$5 kernel size, resulting in 4 MRs being inactive. Lastly, to implement a kernel size of 7$\times$7, a total of 49 MRs are needed for one stride, requiring an entire bank to allocate the necessary resources effectively.
It is worth noting that when implementing a fully connected layer of a DNN, the entire bank is utilized for the calculation. The MAC operations required for the network are divided into 9 operations, which can be executed within each arm by mapping the corresponding weights into the MRs of the arm. This approach efficiently executes the fully connected layer within the OCB architecture.
In total, the OCB consists of 96 banks arranged in an array with 8 columns and 12 rows. Each bank is equipped with 54 MRs, up to 5184 MRs across all banks. This configuration enables a maximum of 5184 MAC operations to be executed in each operational cycle of the OCB.

\begin{figure}[t] 
\centering
\includegraphics [width=0.95\linewidth]{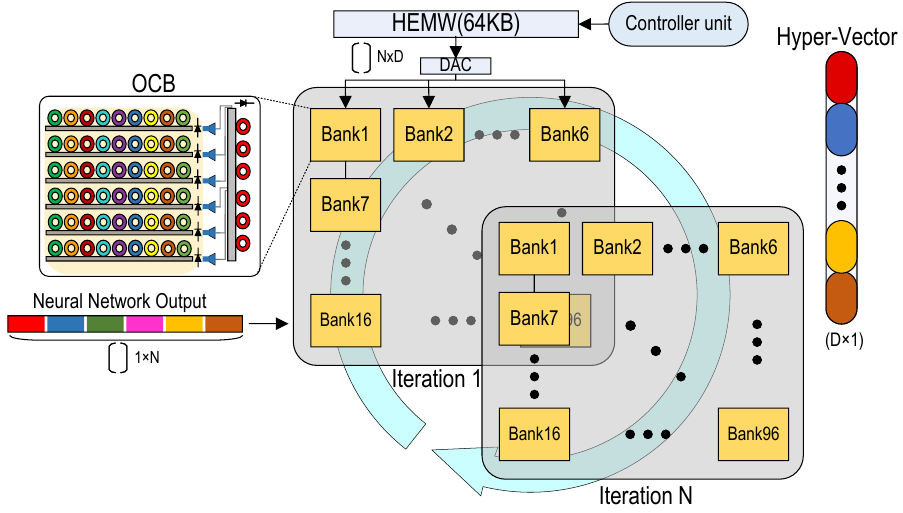}
\vspace{-1.8em}
\caption{Hyper-vector generation in OCB.}
\vspace{-1em}
\label{hdmapping}\vspace{-0.5em}
\end{figure}

\subsection{Photonics-Friendly Encoder}
To achieve the conversion of output data from the DNN output into HV with acceptable efficiency, the encoding HV matrix should have a dimension of 1024. This dimension is adequate to ensure the accuracy discussed in more detail in the experimental results section. The encoding matrix is considered to be $N\times D$, where $N$ represents the maximum output length of a DNN that can be implemented using the OCB, and $D$ denotes the dimensionality of the HV representation. This matrix dimensionality enables efficient mapping and transformation of DNN outputs into HV representations within the constraints and capabilities of the OCB architecture. Similar to the process described in the photonics-friendly MAC engine, mapping the input data to HV involves transferring the values of the encoding HV matrix from its dedicated memory (HEMW). 
The matrix multiplication required for this mapping is performed by executing the MAC operation between the input data and the encoding HV matrix. For clarity, consider the scenario where the output data from the DNN implementation consists of $N$ features. This output is processed by multiplying it row-wise with an encoding matrix of dimensions $N\times D$ to produce a resulting HV of size $1\times D$. It is worth noting that if the number of MAC operations needed to compute the HV exceeds the MAC capacity of the OCB, the HV can be calculated in multiple iterations. In each iteration, the MRs are adjusted with appropriate values from the encoding HV matrix transferred by the HEMW to compute a portion of the HV. After calculating each part, the relevant data from the HEMW updates the MRs to proceed with computing the entire HV, as shown in Fig. \ref{hdmapping}. This iterative approach ensures efficient computation of HV even when exceeding the initial MAC capacity constraints of the OCB.

\vspace{-0.5em}

\section{Experimental Results}
\subsection{Bottom-up Evaluation Framework} 
\begin{figure}[t] 
\centering
\includegraphics [width=0.88\linewidth]{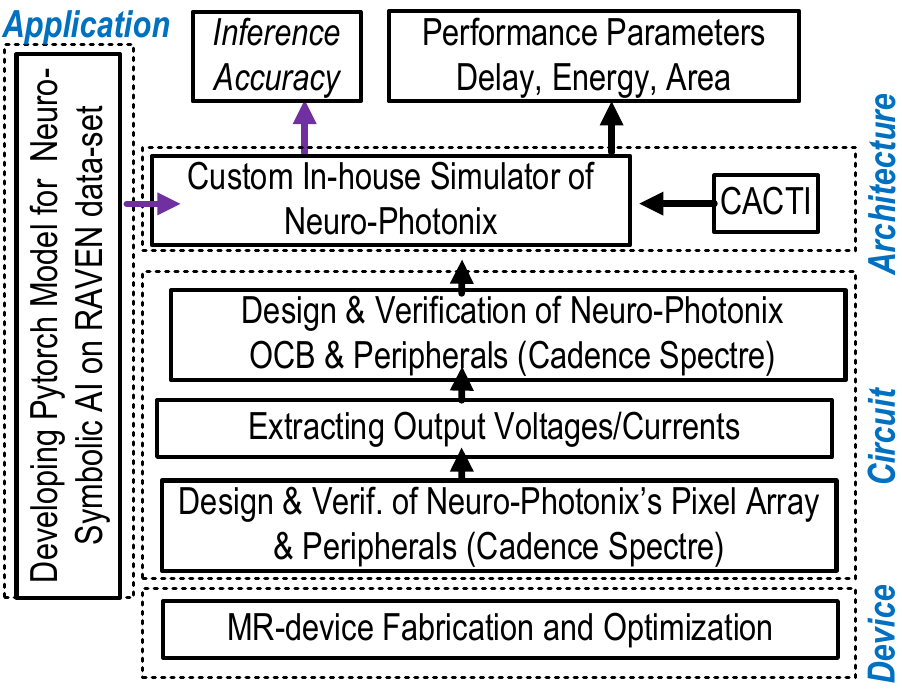}
\vspace{-1.1em}
\caption{Proposed bottom-up evaluation framework.}
\vspace{-1.6em}
\label{framework}
\end{figure}

The framework, depicted in Fig. \ref{framework}, comprises device, circuit, architecture, and application components. Beginning with the device level, we fabricated and optimized MR devices, acquiring circuit parameters for co-simulation with interface CMOS circuits in Cadence Spectre and SPICE. Progressing to the circuit level, we first implement the pixel array and peripheral circuitry using the 45nm NCSU Product Development Kit (PDK) library \cite{NCSU_PDK} in Cadence, from which we derive output voltages and currents. Subsequently, we develop all components excluding kernel banks (implemented in Cacti \cite{thoziyoor2008cacti}) in Cadence Spectre. Our integrated framework combines feature extraction with reasoning at the application level and has been successfully deployed using our proprietary PyTorch AI library. Accordingly, weight parameters are extracted, quantized, fine-tuned, scaled to avoid losing performance, and mapped into the OCB.
At the architecture level, we create an in-house simulator tailored for our platform that computes both execution time and energy consumption necessary for neuro-symbolic AI alongside inference accuracy.

\subsection{Reasoning Task} 
\begin{figure}[b]
    \centering    
\includegraphics[width=0.9\columnwidth]{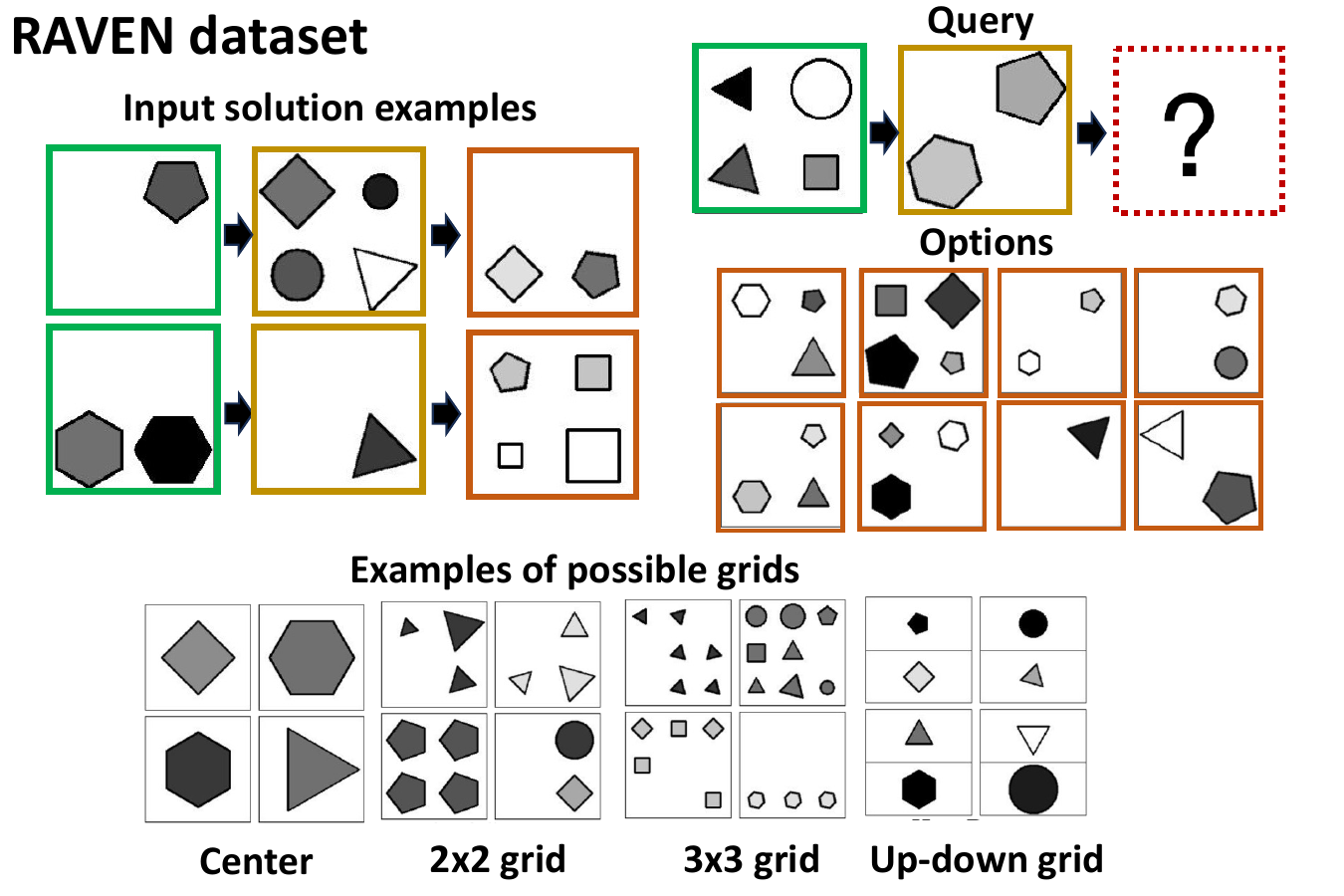}
    \vspace{-4mm}
    \caption{Example from the RAVEN dataset.}
    \label{raven}
    \vspace{-1.4em}
\end{figure}

To rigorously assess the effectiveness of our algorithm, we compared our model with established models referenced in existing research, employing the RAVEN-10000 dataset. The RPM test \cite{raven2019} presents a prime challenge for assessing reasoning due to its combination of visual perception and abstract reasoning. This dataset includes 1,120,000 images and 70,000 instances of RAVEN's Progressive Matrices (RPM), spread evenly across seven distinct figure types. In RPM challenges, as illustrated in Fig. \ref{raven}, participants complete a matrix by recognizing patterns within a series of geometric shapes, which tests their perceptual and logical abilities. 
This extensive dataset serves as a fundamental resource for training and evaluating AI models in reasoning tasks. 
In evaluating our models, we concentrated on several critical metrics: (1) Accuracy, where we calculated the percentage of puzzles correctly solved by each model, providing a direct metric of its effectiveness; (2) Computational efficiency, which examines the inference time and energy consumption per puzzle; (3) Robustness to information loss: We tested each model's ability to withstand noise and data alterations by applying quantization techniques, which is crucial for ensuring model reliability under varied conditions. These measures give a comprehensive overview of how our framework performs across different domains.


\subsection{Accuracy Analysis} 
Our initial evaluation consists of a thorough comparison among several models at full precision and our edge-compatible framework, which includes various architectures such as ResNet+DRT7 \cite{hersche2023neurovectorsymbolic}, DCNet \cite{Zhuo2021DualContrast}, PrAE \cite{Zhang2021ProbabilisticAE}, MRNet \cite{Benny2021ScaleLocalized}, SCL \cite{Wu2020ScatteringCL}, NVSA \cite{hersche2023neurovectorsymbolic}, and Neuro-Photonix based on NVSA with adaptations, and human benchmarks. This set of models is documented in Table~\ref{tab:raven_accs}. For instance, the NVSA model achieves an average accuracy of 98.5\%, setting a benchmark for state-of-the-art reasoning tasks. Focusing on specific model comparisons, the analysis between the high-performing NVSA and our framework, which share similar architectural foundations, offers significant insights. Although this model reached an impressive average accuracy of 98.5\%, our framework was not far behind with an average accuracy of 97.92\%; despite operating at a significantly reduced bit-precision, even for harder tasks such as 3$\times$3, the accuracy drop was not higher than 2\%. We achieved this high performance by strategically implementing 8-bit quantization.

 \begin{table}[b]
\caption{Performance Comparison of NSAI on the RAVEN.} 
\centering
\vspace{-1.2em}
\label{tab:raven_accs}
\scalebox{0.95}{
\begin{tabular}{c|cccc}
\hline
\cellcolor[HTML]{C0C0C0}                                & \multicolumn{4}{c}{\cellcolor[HTML]{C0C0C0}Accuracy}                                        \\ \cline{2-5} 
\multirow{-2}{*}{\cellcolor[HTML]{C0C0C0}Configuration} & Average (\%)    & Centre (\%)     & \multicolumn{1}{l}{3$\times$3 (\%)} & \multicolumn{1}{l}{L-R (\%)} \\ \hline
ResNet+DRT7                                             & 59.6           & 59.6           & 50.4                        & 65.8                        \\
DCNet47                                                 & 93.6           & 97.8           & 86.65                       & 99.8                        \\
PrAE23                                                  & 60.3           & 70.6           & 30.5                        & 88.5                        \\
MRNet10                                                 & 74.7           & 96.2           & 45.9                        & 93.7                        \\
SCL44                                                   & 87.2           & 99.9           & 63.5                        & 96.8                        \\
NVSA                                                    & 98.5           & 99.9           & 96.3                        & 99.9                        \\
\textbf{Our Framework}                                  & \textbf{97.99} & \textbf{99.03} & \textbf{95.15}              & \textbf{99.39}              \\
Human                                                   & 84.4           & 95.5           & 99.39                       & 86.4                        \\ \hline
\end{tabular} \vspace{-5.2 em}}
\end{table} 

\begin{figure}[t]
\begin{center}
 \begin{tabular}{ll}
  \begin{minipage}[c]{0.21\textwidth}
\includegraphics [width=1\linewidth]{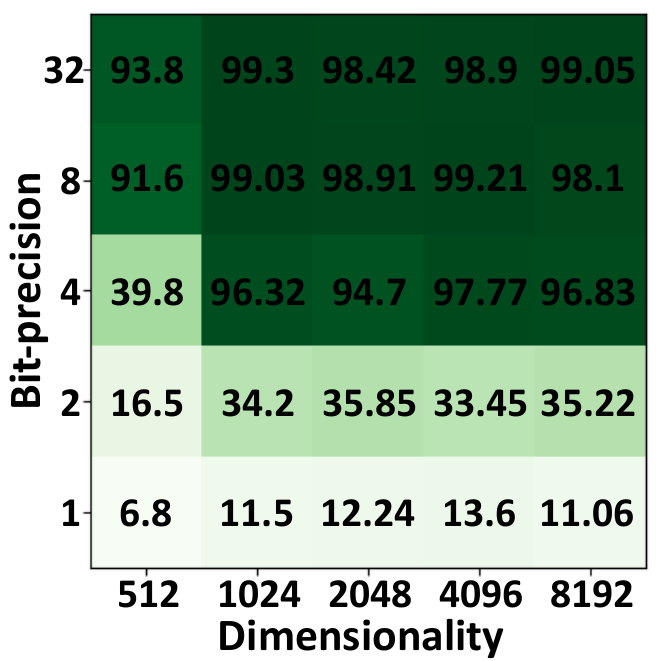} \end{minipage}
 \begin{minipage}[c]{0.2\textwidth}
 \scalebox{0.75}{\begin{tabular}{ccc}
\hline
\textbf{Transfer Cost}                                               & \textbf{Image}                                     & \textbf{Hyper-Vector}                                    \\ \hline
\textbf{Vector Size}                                                 & 16384                                              & 1024                                                     \\
\textbf{\begin{tabular}[c]{@{}c@{}}Data Size \\ (Byte)\end{tabular}}      & 65536                                              & 512                                                     \\
\textbf{\begin{tabular}[c]{@{}c@{}}Energy \\ (mJ)\end{tabular}}      & 7680                                           & 60                                                  \\ \hline
\textbf{\begin{tabular}[c]{@{}c@{}}BLE Energy\\ Efficiency\end{tabular}} & \multicolumn{2}{c}{\begin{tabular}[c]{@{}c@{}}15mW/1Mb \\ \cite{decuir2010blacktooth}\end{tabular}} \\ \hline
\end{tabular}} \\ 
 \end{minipage} &
 \\
 \hspace{2.1 cm}   \small (a) & \hspace{-2.4 cm}  \small (b)\\\vspace{-1em}
 \end{tabular} \vspace{-1em}
\caption{(a) Heatmap displaying accuracy across various dimensions and quantization levels for the CENTER RAVEN, (b) Data transfer cost of traditional reasoning at the cloud vs. Neuro-Photonix generated HVs.} \vspace{-2em}
\label{fig:heatmap}
\end{center}
\end{figure}
Next, we explore the efficacy of Neuro-Photonix relative to standard full-precision benchmarks by examining average accuracy across a range of quantization scenarios for the neural block and the impact on dimensionality for the HDC block of the network. The findings are illustrated in Fig.~\ref{fig:heatmap}(a), which presents a heatmap detailing the accuracy of the center configuration of the RAVEN over various dimensions (512, 1024, 2048, 8196) and bit precisions (2, 4, 8, 32, or full-precision). With standard 32-bit precision, without any quantization, the minimum dimensionality required to achieve near state-of-the-art performance is 1024. 
A reduction to a 512 dimension results in a substantial 6\% decrease in accuracy at full precision. When precision is reduced to 8 bits, the accuracy of our framework slightly diminishes to 98.1\% for dimensions greater than 1024, demonstrating the model's resilience to reduced precision. At this bit level, the impact of dimensionality becomes less pronounced, except at the 512 level, where the accuracy drops by 2.2\% from the full-precision case and by almost 8\% from state-of-the-art. This phenomenon is highly tied to the mathematical underpinnings of HDC and its connection to Random Fourier Features \cite{randomfourierfeatures}, where, depending on the complexity of the task, there is a requisite minimum dimensionality for adequate kernel approximation, which in this scenario it needs to be higher than 512. As the complexity of the tasks increases, higher dimensions are necessary to improve kernel approximation, which in turn brings higher performance.

At a quantization of 4 bits, we identified a critical trade-off for our framework at the 1024 dimensionality. Here, there is a loss of approximately 3\% in accuracy regarding reasoning tasks, though this loss is not as marked as at the 512 dimensionality. For dimensions beyond 1024, no improvements in accuracy are observed, indicating a critical threshold beyond which performance degradation becomes noticeable. 
Overall, for bit-precisions lower than 4 bits, we observe that the quality diminishes too much for it to be usable, given the complexity and difficulty of the RAVEN reasoning dataset, which highlights the intricate balance required between model efficiency and performance.
This evaluation serves as a foundational step in understanding and optimizing the trade-offs inherent in our design of the hardware architecture of Neuro-Photonix.

\subsection{Data Transmission Cost} As shown in Fig.~\ref{fig:heatmap}(b), to evaluate the benefits of our framework, we estimate the transmission cost to the cloud, which is often overlooked but is particularly significant. We use blacktooth Low Energy (BLE) 4.0 \cite{decuir2010blacktooth} as the communication protocol. In the case of traditional reasoning at the cloud, the transmission cost over blacktooth is significant since the entire input image obtained after sensing has to be transmitted to the cloud. In contrast, even though the design has to bear the overhead of encoding near-sensor, the size of the packet being transmitted is drastically reduced, diminishing transmission costs to cloud infrastructure by 128$\times$, and minimizing the overall system-level cost of reasoning, enabling near-sensor integration.

\subsection{Energy Consumption and Performance}
To evaluate the energy consumption and performance of  Neuro-Photonix, we consider the ResNet18 network. During evaluations, we realized aside from conversion energy (ADCs and DACs), the tuning of MRs also accounts for a significant portion of the architecture's energy consumption, warranting consideration. Therefore, two evaluation methods are studied. 
In the first method, we assume that each time new activations are applied to the OCB. Then, the inputs are rearranged, and the MRs are retuned with weight values, regardless of whether the new activations require multiplication by new weights or not. This method is referred to as the Non-Re-Using Weights method (NRU). In the second method, we refrain from tuning new weights on the OCB until all activations intended to be multiplied by those weights are applied. Subsequently, we retune and rearrange the weights before applying their corresponding activations. The second method enables substantial savings in weight conversion energy consumed by DACs and MRs tuning energy.

\begin{figure}[t] 
\centering
\includegraphics [width=1\linewidth]{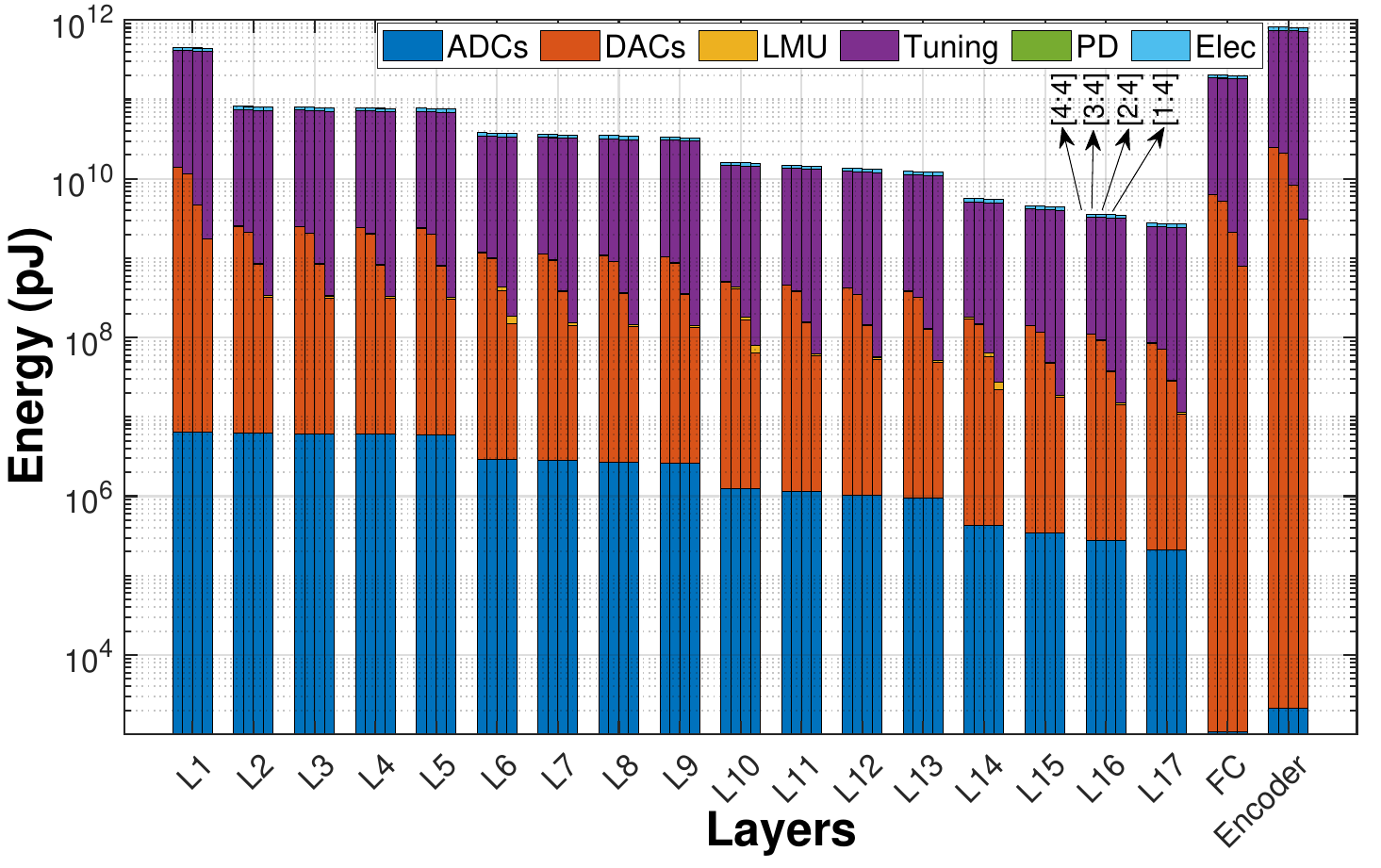}
\vspace{-1.7em}
\caption{Break-down of energy consumption for ResNET18 layers and encoder layer under different [W:A] in NRU.}
\label{energyNRU}
\end{figure}

Fig. \ref{energyNRU} provides a breakdown of energy consumption components by layer for four weight and activation [W:A] configurations of the ResNET18 and encoder layer mapped on Neuro-Photonix in the NRU method. \textcolor{black}{The [W:A] ratio refers to the ratio of Weight precision to Activation precision during the inference phase. This ratio is critical as it directly impacts the computational and memory requirements of the system.}
We observe that (1) The utilization of CBC and LDU components has resulted in a reduction of ADCs and DACs conversions for converting input image data and input activations from the previous layer, making the ADCs and DACs energy components comparable to other energy components; (2) ADCs, DACs, and tuning constitute the most significant portions of energy consumption; (3) As the number of weights in the fully connected layer of the neural dynamic and the encoder layer of the symbolic AI is larger than that of in the conv. layers, it resulted in higher DACs and tuning energies compared to the ADCs energy in these two layers; (4) Reducing the bit-width of the weights has only an insignificant impact (less than 1\%) on overall energy consumption reduction.

\begin{figure}[b]
\centering
\includegraphics [width=1\linewidth]{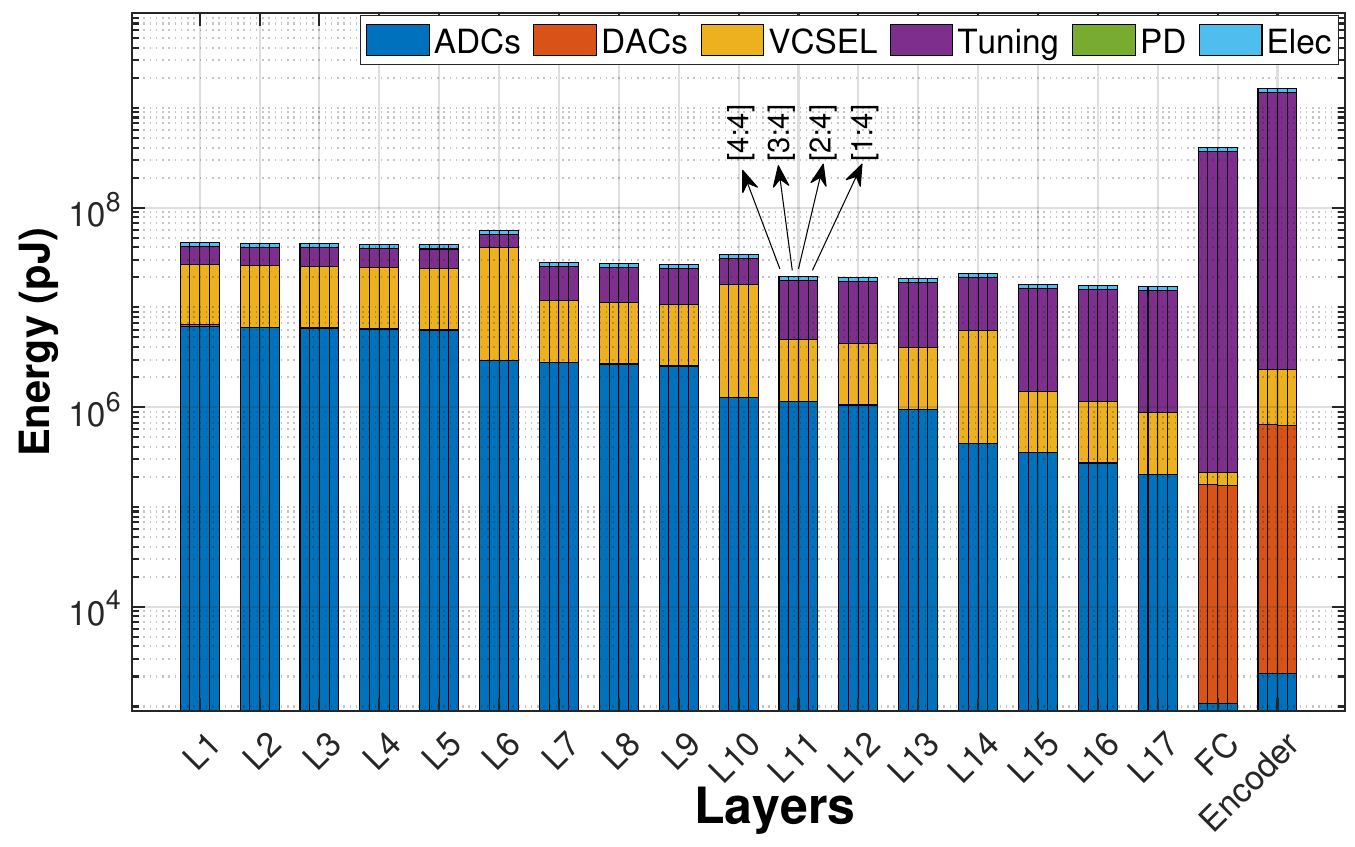}
\vspace{-1.2em}
\caption{Break-down of energy consumption for ResNET18 layers and encoder layer under different [W:A] in RU.}
\vspace{-1.6em}
\label{energyRU}
\end{figure}

Fig. \ref{energyRU} shows the breakdown of energy consumption utilizing the RU method. As depicted, utilization of the RU method has resulted in a significant reduction in the energy consumption of tuning and DACs, leading to a notable $\sim$800$\times$ and $\sim$500$\times$ reduction in the total energy required for processing the neural dynamics and symbolic AI, respectively. 
However, such reduction in energy consumption is smaller in the symbolic AI part as evident from Fig. \ref{pie}(a) and (c), primarily due to the more significant number of weight parameters in the encoding layer, necessitating more DACs and tuning energy.

\begin{figure}[b] 
\centering
\includegraphics [width=1\linewidth]{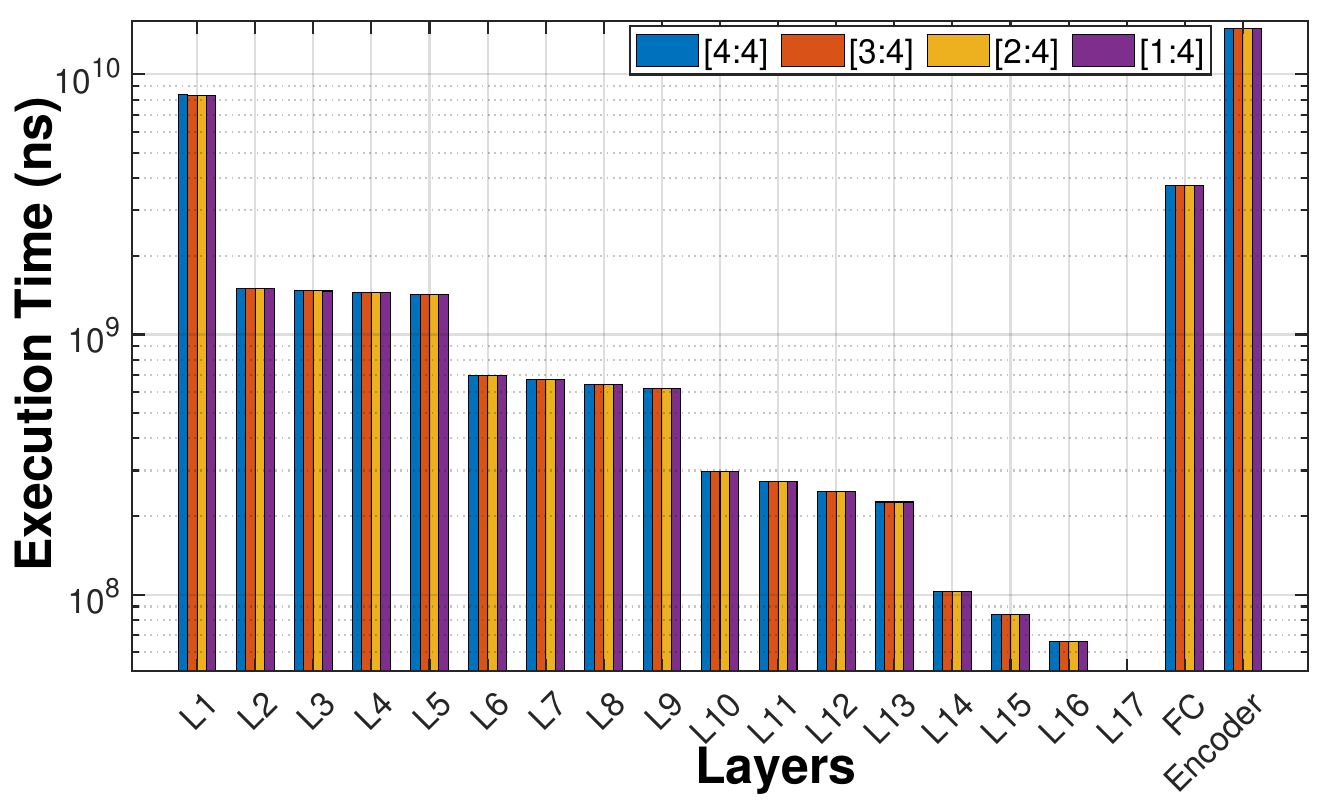}
\vspace{-2em}
\caption{Execution time of ResNET18 layers and encoder layer under different [W:A] settings in NRU method.}
\vspace{-1.9em}
\label{latencyNRU}
\end{figure}
\begin{figure} 
\centering
\includegraphics [width=1\linewidth]{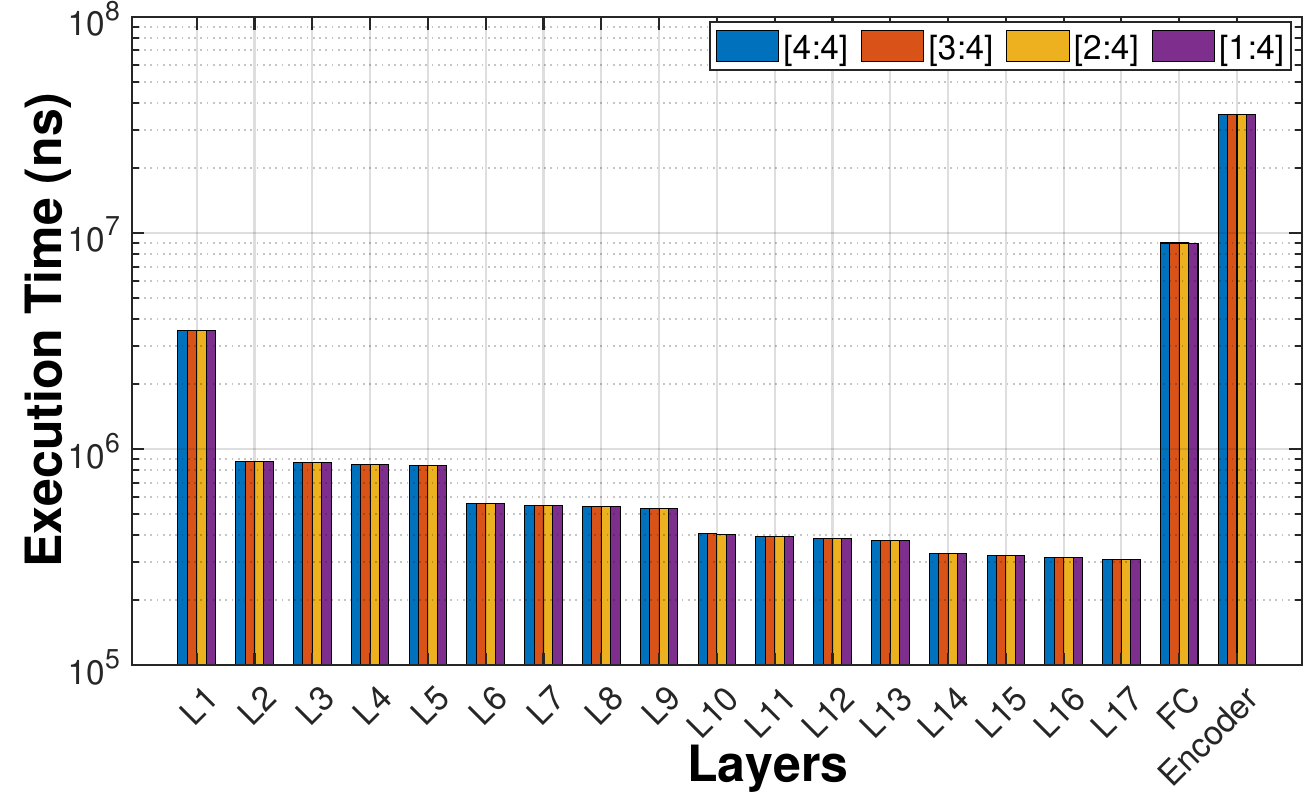}
\vspace{-2em}
\caption{Execution time of ResNET18 layers and encoder layer under different [W:A] settings in RU method.}
\vspace{-1.0em}
\label{latencyRU}
\end{figure}
\begin{figure}[b] 
\centering
\includegraphics [width=0.90\linewidth, height=2.3cm]{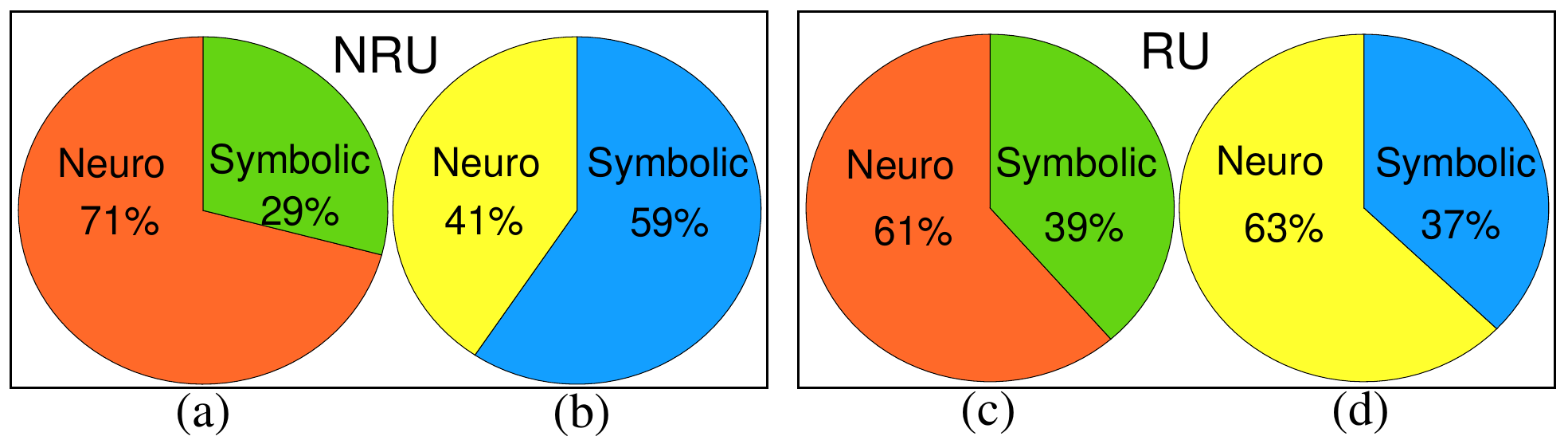}
\vspace{-1.2em}
\caption{Distribution of energy consumption (a) and (c) and
execution time (b) and (d) for neuro-symbolic AI task.}
\vspace{-1.2em}
\label{pie}
\end{figure}

Layer-wise execution time of the ResNET18 layers and encoder layer using Neuro-Photonix architecture in NRU method under various [W:A] is displayed in Fig. \ref{latencyNRU}. The encoder layer shows the largest execution time. In processing layers, tuning the MRs is the most time-consuming part. Consequently, with its higher number of weights, the encoder layer experiences a more significant delay than the other layers. Also, changing the bit-width of the weight has a minimal effect (less than 1\%) on the execution time of the layers. Thus, reducing the bit-width and, consequently, the accuracy does not lead to a significant performance improvement. 
Fig. \ref{latencyRU} shows the layer-wise execution time of ResNET18 in the RU setup. Utilizing the RU method has resulted in significant savings in the execution time of both the neural dynamics and symbolic AI parts due to the reduction in MR's tuning time. Compared to the NRU method, the RU method has reduced the required processing time of the neural dynamics layers by $\sim$400$\times$ and the symbolic AI layer by $\sim$1000$\times$. It is evident from Fig. \ref{pie}(b) and (d) that the symbolic AI part has benefited the most from the RU method.
As an example in the setting of [3:4], the total required energy for processing the layers of ResNET18 and the encoder layer is 2796 mJ for the NRU method and 4.1 mJ for the RU method. The processing time of the Neuro-Photonix working in the NRU and RU methods is 36.9 s and 56.4 ms, respectively. The pie charts in Fig. \ref{pie} display the distribution of energy consumption and execution time between the neuro and symbolic parts for the [3:4] configuration. We observed a significant change in the share of the Symbolic part's total execution time when transitioning from the NRU to the RU processing method, decreasing from 59\% to 37\%. This shift is primarily due to the tuning operation, the most time-consuming process in our processing core, which is far surpassing other component delays.\vspace{-1em}


\begin{table*}[b]
\caption{Performance comparison with optical designs.}\vspace{-1em}
\centering
\scalebox{1}{
\begin{tabular}{lcccccc}
\hline
\rowcolor[HTML]{C0C0C0} 
\multicolumn{1}{c}{\cellcolor[HTML]{C0C0C0}}                                                                                           & \cellcolor[HTML]{C0C0C0}                                                                                       & \cellcolor[HTML]{C0C0C0}                                                                                    & \cellcolor[HTML]{C0C0C0}                                  & \multicolumn{3}{c}{\cellcolor[HTML]{C0C0C0}\textbf{Accuracy (\%)}}                                \\ \cline{5-7} 
\rowcolor[HTML]{C0C0C0} 
\multicolumn{1}{c}{\multirow{-2}{*}{\cellcolor[HTML]{C0C0C0}\textbf{\begin{tabular}[c]{@{}c@{}}Designs \&\\ $[$W: A$]$\end{tabular}}}} & \multirow{-2}{*}{\cellcolor[HTML]{C0C0C0}\textbf{\begin{tabular}[c]{@{}c@{}}Process node\\ (nm)\end{tabular}}} & \multirow{-2}{*}{\cellcolor[HTML]{C0C0C0}\textbf{\begin{tabular}[c]{@{}c@{}}Max Power \\ (W)\end{tabular}}} & \multirow{-2}{*}{\cellcolor[HTML]{C0C0C0}\textbf{KFPS/W}} & \multicolumn{1}{l}{\cellcolor[HTML]{C0C0C0}\textbf{MNIST}} & \textbf{CIFAR10} & \textbf{CIFAR100} \\ \hline
baseline [32:32]$^\S$                                                                                                                  & 8                                                                                                              & 200                                                                                                         & -                                                         & 98.53                                                      & 90.46            & 67.8              \\
LightBulb [1:1] \cite{zokaee2020lightbulb}                                                                                             & 32                                                                                                             & 68.3                                                                                                        & 57.75                                                     & 96.7                                                       & -                & -                 \\
HolyLight [4:4] \cite{liu2019holylight}                                                                                                & 32                                                                                                             & 66.9                                                                                                        & 3.3                                                       & 98.9                                                       & 88.5             & -                 \\
HQNNA \cite{sunny2022silicon}                                                                                                          & 45                                                                                                             & -                                                                                                           & 34.6                                                      & -                                                          & 89.68            & 61.95             \\
Robin [1:4] \cite{sunny2021robin}                                                                                                      & 45                                                                                                             & 106                                                                                                         & 46.5                                                      & -                                                          & 62.5             & 45.6              \\
CrossLight [4:4] \cite{sunny2021crosslight}                                                                                            & -$^{*}$                                                                                                        & 84-390                                                                                                      & 10.78-52.59                                               & 92.6                                                       & 78.85            & -                 \\
\textbf{Neuro-Photonix [4:4]}                                                                                                               & \textbf{45}                                                                                                    & \textbf{5.28}                                                                                               & \textbf{61.61}                                            & \textbf{98.12}                                             & \textbf{88.87}   & \textbf{64.22}    \\
\textbf{Neuro-Photonix [3:4]}                                                                                                               & \textbf{45}                                                                                                    & \textbf{2.71}                                                                                               & \textbf{117.65}                                           & \textbf{98.05}                                             & \textbf{86.3}    & \textbf{61.04}    \\
\textbf{Neuro-Photonix [2:4]}                                                                                                               & \textbf{45}                                                                                                    & \textbf{1.46}                                                                                               & \textbf{188.24}                                           & \textbf{93.95}                                             & \textbf{70.55}   & \textbf{41.4}     \\
 \hline
\end{tabular}}

\small$^{\S}$NVIDIA Geforce RTX 3060Ti GPU.
$^{*}$Data is not reported/not achievable in the paper. 
\label{results}
\end{table*}

\subsection{Comparison}
\subsubsection{Comparison with Electronic Accelerators}
Given the pioneering nature of Neuro-Photonix as the first-of-its-kind near-sensor neuro-symbolic AI accelerator, direct performance comparisons with other designs prove challenging. Nevertheless, to shed light on its capabilities, we undertake a relative assessment of the neural dynamic component.
We compare Neuro-Photonix power consumption with three prominent digital electronic accelerators, i.e., Eyeriss \cite{chen2017eyeriss}, YodaNN \cite{andri2018yodann}, and AppCip \cite{tabrizchi2023appcip}, each with distinct parallelism techniques and hardware mappings. Eyeriss utilizes a spatial architecture for energy-efficient processing, YodaNN is optimized for binary-weight CNNs, and AppCip implements parallel analog convolution-in-pixel. Our cross-layer simulation results demonstrate that the Neuro-Photonix significantly outperforms electronic accelerators for processing ResNET18. Specifically, Neuro-Photonix reduces power consumption by factors of 19$\times$, 28$\times$, and 17.6$\times$ compared to Eyeriss \cite{chen2017eyeriss}, YodaNN \cite{andri2018yodann}, and AppCip \cite{tabrizchi2023appcip}, respectively.

\subsubsection{Comparison with Optical Accelerators.} 
Table \ref{results} presents our extensive simulation results for selected MR-based optical accelerators and Neuro-Photonix across various [W:A] configurations, compared with the baseline, an NVIDIA Geforce RTX 3060Ti GPU. The DNN accelerators under test include LightBulb \cite{zokaee2020lightbulb}, HolyLight \cite{liu2019holylight}, HQNNA \cite{sunny2022silicon}, Robin \cite{sunny2021robin}, and CrossLight \cite{sunny2021crosslight}, as discussed in the background section. To ensure an unbiased evaluation, we built the designs from the ground up to closely match the original designs, using our evaluation framework and in-house simulator. The results are reported within a reasonable area constraint for all accelerators ($\sim$20-60$mm^2$). 
Taking into account the minimum phase cross-talk ratio in our architecture, we estimate the area of our proposed optical core to be  ($\sim$5$mm^2$) \cite{sunny2021robin}.

Here we list our key observations. 
$(1)$ We observe that the variants of Neuro-Photonix exhibit exceptional power efficiency compared to other designs on the VGG9 model running CIFAR100. For instance, Neuro-Photonix [3:4] consumes only 2.71 W, which can be supported by the low power budget of edge devices. In contrast, the best low-power accelerator, HolyLight \cite{liu2019holylight}, requires 66.9 W or more \cite{sunny2021robin}. This significant power efficiency is achieved by $(i)$ eliminating the need for MRs tuned by activation parameters, thereby saving the tuning power required for the MRs, and $(ii)$ reducing the additional power and area requirements due to the extensive use of ADCs and DACs. 
$(2)$ On average Neuro-Photonix reduces power consumption by $\sim$73$\times$, 24.68$\times$, 30.9$\times$ compared with the baseline [32:32],  HolyLight [4:4] \cite{liu2019holylight}, and CrossLight [4:4] \cite{sunny2021crosslight}, respectively.
$(3)$ As we reduce the weight bit-width, the power consumption can be reduced at the cost of accuracy degradation, where Neuro-Photonix [3:4] achieves $\sim$2$\times$ power saving at the cost of 3.17\% accuracy drop.
$(4)$ Regarding throughput ($\frac{frame}{second}$) per watt, Neuro-Photonix [3:4] achieves 117.65 kilo FPS/W, doubling the inference performance compared to the best result reported for LightBulb \cite{zokaee2020lightbulb}. Overall, taking into account the test accuracy results across three datasets, Neuro-Photonix-MX [4:4][3:4] provides the best performance-quality ratio with 84.4 kilo FPS/W. 

$(5)$ Our experiments show that Neuro-Photonix with [3:4] and [4:4] configurations maintain good accuracy across three datasets. Neuro-Photonix [4:4] achieves the second-highest accuracy on MNIST and CIFAR10, following HolyLight \cite{liu2019holylight} and HQNNA \cite{sunny2022silicon}, respectively, while offering higher KFPS/W. $(6)$ We found that accuracy decreases with bit-width sensitivity in activations and weights. The main accuracy drop in Neuro-Photonix is due to the ADC-less imager affecting the first layer, which can be improved with hardware noise-aware training.

\vspace{-0.7em}
\section{Conclusions}
This work introduces Neuro-Photonix as the first-of-its-kind near-sensor photonic accelerator for neural symbolic AI computing to realistically perform reasoning tasks near the sensor.
In this architecture, dynamic neural implementation leverages the potential efficiency offered by photonic devices. The proposed architecture facilitates the MAC operation on analog input data, essential for various neural network layers implementation, in a single cycle. Additionally, this design supports HyperDimensional (HD) computing, enhancing the neural network's accuracy, particularly with low-precision input data. Furthermore, this approach minimizes overall power consumption, conversion latency, and processing time within both established cloud-centric architectures and newly designed accelerators. Simulation results demonstrate that Neuro-Photonix achieves 30 GOPS/W and reduces power consumption by a factor of 20.8 and 4.1 on average compared to recent ASIC baselines and photonic accelerators while preserving accuracy.
For future work, we suggest further development of the Neuro-Photonix architecture to incorporate the similarity evaluation needed in HDC directly within the architecture, eliminating the traditional reliance on cloud-based processing. Additionally, the proposed design's potential for implementing MAC operations could be further leveraged by enabling approximate computing within the photonic framework.

\ifCLASSOPTIONcaptionsoff
  \newpage
\fi

\bibliographystyle{IEEEtran}
\bibliography{IEEEabrv,./main}\vspace{-2em}




\end{document}